\documentclass[aps,prl,a4paper,superscriptaddress,
                 twocolumn,showpacs]{revtex4}
\usepackage[dvips]{graphicx} \usepackage{graphics}

\begin{document}
\title{Why Are Alkali Halide Solid Surfaces Not Wetted By Their Own Melt?}

\author{T. Zykova-Timan}
   \affiliation{International School for Advanced Studies (SISSA), and
   INFM Democritos National Simulation Center, Via Beirut 2-4,
   I-34014 Trieste, Italy}

\author{D. Ceresoli}
   \affiliation{International School for Advanced Studies (SISSA), and
   INFM Democritos National Simulation Center, Via Beirut 2-4,
   I-34014 Trieste, Italy}

\author{U. Tartaglino}
   \affiliation{International School for Advanced Studies (SISSA), and
   INFM Democritos National Simulation Center, Via Beirut 2-4,
   I-34014 Trieste, Italy}
   \affiliation{IFF, FZ-J\"ulich, 52425 J\"ulich, Germany}

\author{E. Tosatti}
   \affiliation{International School for Advanced Studies (SISSA), and
   INFM Democritos National Simulation Center, Via Beirut 2-4,
   I-34014 Trieste, Italy}
   \affiliation{International Centre for Theoretical Physics (ICTP),
   P.O.Box 586, I-34014 Trieste, Italy}

\date{\today}

\begin{abstract}
Alkali halide (100) crystal surfaces are anomalous, being very poorly wetted by their
own melt at the triple point. We present extensive simulations for NaCl,
followed by calculations of the solid-vapor, solid-liquid and liquid-vapor
free energies showing that solid NaCl(100) is a non-melting surface, and that
its full behavior can quantitatively be accounted for within a simple (BMHFT)
model potential. The incomplete wetting is traced  to the conspiracy
of three factors, namely: surface anharmonicities stabilizing the solid
surface; a large density jump causing bad liquid-solid adhesion; incipient
NaCl molecular correlations destabilizing the liquid surface. The latter
is pursued in detail, and it is shown that surface short range charge
order acts to raise the surface tension because incipient NaCl molecular
formation anomalously reduce the surface entropy of liquid NaCl much
below that of solid NaCl(100).
\pacs{63.03.Cd; 68.08.Bc; 68.08.De}

\end{abstract} \maketitle
Molten alkali halides and their surfaces have long
been studied experimentally~\cite{janz,croxton} and
theoretically~\cite{croxton,goodisman,evans,heyes,aguado,rpm}. 
Much less attention
has been devoted to solid alkali halide surfaces at high temperatures,
and especially to their wetting habit at the melting point. Yet,
these crystal surfaces behave anomalously in that respect. Whereas most
liquids would wet their own solid-vapor interfaces -- the solid surfaces
undergoing surface melting at the melting point $T_m$ (triple point
wetting~\cite{dietrich})--  molten salts wet their own solid surface
only incompletely. The external and internal contact angles $\theta$
and $\phi$ of a partially wetting liquid droplet onto its own solid are 
connected to the free energies $\gamma$ of the solid-vapor (SV), solid-liquid
(SL) and liquid-vapor interfaces by Young's equation~\cite{nozieres}:
\begin{equation}\label{eq:young}
   \gamma_{SV} = \gamma_{SL}\,\cos\phi + \gamma_{LV}\,\cos\theta.
\end{equation}
Incomplete, or partial, triple point wetting, with
$\phi \sim 0$ but with $\theta > 0$ and stable near $T_m$, implies that some
physical mechanism must be at work making $\gamma_{SV} < \gamma_{SL} +
\gamma_{LV}$. For liquid NaCl on NaCl(100) at $T_m =$ 1074~K, a partial
wetting angle $\theta \sim 48^\circ$ is well established , with similar
results holding for other alkali halides too.~\cite{mutaftschiev,komunjer}
This angle is exceptionally large -- and thus the corresponding
self-wetting at the triple point exceptionally poor -- even when compared
with strongly non-wetting metal surfaces, where e.g. $\theta \sim 15^\circ$
measured for liquid Pb/Pb(111)~\cite{frenken}, or $\sim 18^\circ$ obtained
by simulation of liquid Al/Al(111)~\cite{ditolla}. Should liquid surface
layering~\cite{rice} be, as in the case of metals,~\cite{ditolla, report} be the
culprit for the incomplete wetting of alkali halides too, the layering
magnitude and its effects should be exceptionally strong.
However, all of the existing molten salt theory and simulations indicate
the opposite, namely a soft, smooth, layering-free liquid-vapor
interface~\cite{heyes,aguado,rpm}. That leaves the partial wetting wide open
for another explanation. In a more general context, it seems desirable
to pursue a case study of solid-liquid-vapor coexistence at the triple
point in a simple but realistic model that could be addressed at the fully
microscopic level. 
We wish in particular to understand what may control wetting and
adhesion in a specific and chemically unambiguous case such as that
of a liquid with its own solid. For these reasons and in order to shed light on the underlying
physics, we undertook extensive simulations of the NaCl(100) solid
surface, of the liquid NaCl surface, and of the solid-liquid interface
around $T_m$.

NaCl, our prototype alkali halide, was described by the classic
Born-Mayer-Huggins-Fumi-Tosi (BMHFT) two body potential~\cite{fumi}.
Polarization forces, though not negligible~\cite{aguado}, were sacrificed
in the present context, where computational simplicity is essential to
reach a unified description of all possible interfaces with very large
sizes and simulation times.
Eventually, as it turnes out, quantitative accuracy in the description
of NaCl interfaces seems anyway quite good in the BMHFT model.

Bulk systems were first simulated by molecular dynamics (MD)
at constant volume with cubic simulation cells comprising up to about 5000
NaCl units. Surfaces were studied with periodically repeated slabs --
consisting of 12$\div$24 planes with 64 NaCl units each -- separated by
100$\div$120 \AA\ of vacuum. We treated long range forces in full using a
3D Ewald summation. Despite the size and time limitations imposed by long
range forces, great care was taken to run simulations long enough for
a clear equilibration, typically 100-300 ps at $T_m$. In preliminary
bulk simulations, we determined the equilibrium lattice spacing
$a(T)$ as a function of temperature at zero pressure. All subsequent
simulations were then performed at fixed cell size, by enforcing $a(T)$ at
each temperature.
The calculated room temperature $a(T_R)$ and linear expansion
coefficient, 5.683~\AA\ and 40.5 $\cdot$ 10$^{-6}$~K$^{-1}$
are in very good agreement with experiment, 5.635~\AA\ and 38.3 $\cdot$
10$^{-6}$~K$^{-1}$.
The bulk melting temperature $T_m$ -- practically identical to the
triple point temperature owing to Clapeyron's equation -- was established 
by simulation of liquid-solid
coexistence at zero pressure and found to be 1066$\pm$20~K
(1074~K experimental).\cite{SL}
This value is also in perfect agreement with 1064$\pm$14~K independently
obtained for the BMHFT potential by Anwar \emph{et al}~\cite{frenkel}. Our
theoretical volume expansion and latent heat of melting are 27\% and
0.2899  (experimental 26\% and 0.2915 eV/molecule~\cite{janz}). 
Full details will be presented in a forthcoming paper~\cite{long}.

In surface MD simulations, defect free NaCl(100) was first of
all found to remain indefinitely solid and totally dry for all
temperatures up to $T_m$, and for even the longest simulation times
$\sim$ 1 ns. Moreover, in a metastable state, simulated NaCl(100)
remained crystalline even above $T_m$. By melting initially a few
surface layers, and then observing in a subsequent canonical MD run
that these liquid layers spontaneously recrystallized even above
$T_m$,~\cite{carnevali}, we established the existence of a multi-monolayer 
thick nucleation barrier against surface melting up to about 1115~K~$\simeq T_m +$~50~K, then
a thin monoatomic nucleation barrier until a 1215~K~$\simeq T_s = T_m +$~150~K.
Only at $T_s$, well above $T_m$, does solid NaCl(100) become
locally unstable and spontaneously melt. This metastable behavior was
found to persist even in presence of surface defects, such as molecular
vacancies~\cite{note1} or steps, at least up to 1115~K~\cite{long}.
Altogether, these results characterize NaCl(100) as a clear case of
surface non-melting,~\cite{carnevali,frenken,report} a prediction that
deserves to be tested in experiment. For a short enough time, it should
be possible for example to overheat NaCl(100) and other alkali halide
surfaces substantially above $T_m$ without melting them.

\begin{figure}\begin{center}
    \includegraphics[width=0.45\textwidth]{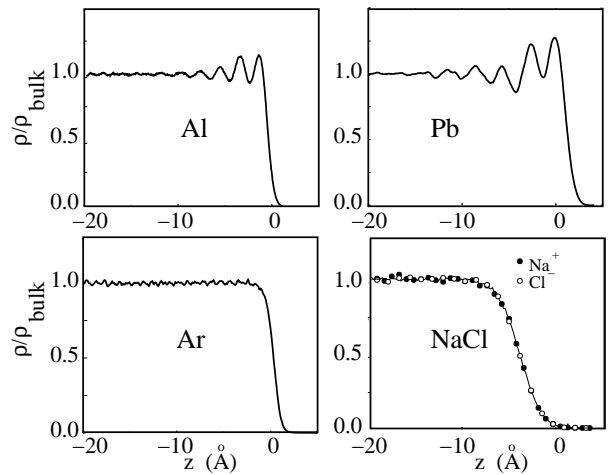}
    \caption{Simulated density profile of liquid surfaces~\cite{report} of (a) Pb,
  (b) Al, (c) LJ\,(Ar) in comparison with (d) $\rho(Na)$
    and  $\rho(Cl)$ profiles in NaCl at $T_m$; solid line: average of the two.}
    \label{profiles}
\end{center}\end{figure}

\begin{figure}\begin{center}
    \includegraphics[width=0.45\textwidth]{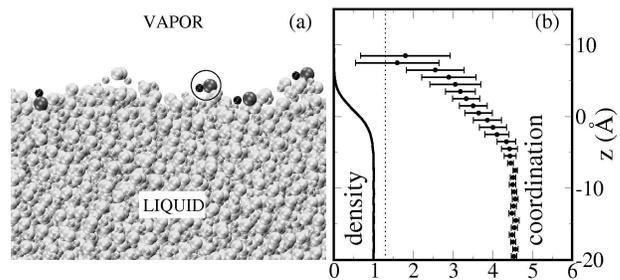}
    \caption{a) Simulation snapshot of the NaCl liquid surface at $T_m$.
    Notice the large thermal fluctuations, and some nearly molecular
    configurations highlighted in the outermost region;
    b) Coordination number $N(Z)$ and density profile showing the smooth
    crossover from liquid ($N$ = 4.6) to molecular vapor
    ($N$ = 1.3 dotted line). Density profile as in Fig.~\ref{profiles} in
    units $\rho/\rho_\mathrm{bulk}$.}
    \label{liq}
\end{center}\end{figure}

We proceeded next to simulate the liquid NaCl surface, more precisely
melted NaCl slabs. The (x,y) averaged liquid local density profile
$\rho(z)$ (Fig.~\ref{profiles}d) is confirmed to be remarkably smooth as
found in previous studies~\cite{heyes,aguado,rpm}, devoid of layering 
unlike Al, Pb (Fig.~\ref{profiles}a,b),
and much more diffuse (Fig.~\ref{profiles}d) than even that of liquid Ar
(Fig.~\ref{profiles}c).
The nature of diffuseness of the NaCl
surface is demonstrated by the simulation snapshot of Fig.~\ref{liq}a,
showing very pronounced fluctuations~\cite{note2}, in the instantaneous
surface profile. This picture is suggestive of a low surface tension, high
entropy surface, in apparent contradiction with the massive non-wetting
of solid NaCl(100) by its melt, which is only favored by a sufficiently large
$\gamma_{LV}$ (Eq.~(\ref{eq:young})).
In order to clarify the situation, we undertook direct calculations
of interface free energies $\gamma_{SV}$, $\gamma_{LV}$, and
$\gamma_{SL}$. We obtained $\gamma_{LV}$ as the surface stress of the
simulated liquid slab (two interfaces) through the Kirkwood-Buff formula:
\begin{eqnarray}\label{KBnew}
\gamma_{LV} &=&
\frac{1}{2}\int_{0}^{L_{z}} dZ[\sigma_{\parallel}(Z) - \sigma_{\perp}(Z)] {}
\nonumber \\
&=& -\frac{1}{8}\int_{-\infty}^{\infty} dZ \int
d^{3}\textbf{r}_{ij} \sum_{i\alpha,j\beta} 
\frac{x^{2}_{ij}+y^{2}_{ij}- 2z^2_{ij}}{r_{ij}} \big[\delta_{\alpha \beta} + {}
\nonumber \\
&+& \lambda(1 - \delta_{\alpha \beta})\big]
f_{\alpha\beta}(r_{ij}) g_{\alpha \beta}^{(2)} (\textbf{r}_{ij};
Z)\rho_{\alpha}(Z)\rho_{\beta}(Z){} \nonumber \\
&=&-\frac{1}{8L_{x}L_{y}}\langle \sum_{i\alpha,j\beta}
\frac{x^{2}_{ij}+y^{2}_{ij}-2z^2_{ij}}{r_{ij}}
\big[\delta_{\alpha \beta} + {} \nonumber \\
&+& \lambda (1-\delta_{\alpha \beta})\big] f_{\alpha\beta}(r_{ij}) \rangle
\end{eqnarray}
where: $(\alpha,\beta) = (+,-)$, $i\alpha$ and $j\beta$ denote ions at site $i$ or $j$, $Z$ is the distance normal to the interface,
$L_{x}, L_{y}$ are the $(x,y)$ dimensions of the supercell
and $\sigma_{\parallel}=\frac{1}{2}(\sigma_{xx}+\sigma_{yy})$ and
$\sigma_{\perp}=\sigma_{zz}$
are the tangential and normal components of the stress tensor respectively.
Here $\langle \ \rangle$ denotes a canonical average and $\sum_{i,j}$ is over
all pairs of particles. Moreover ${\textbf{r}}_{ij}= (x_{ij}, y_{ij}, z_{ij}$)
is the interatomic distance, $f_{\alpha \beta}(r_{ij})$ is the force between
atoms $i$ and $j$, $g_{\alpha \beta} ({\textbf{r}}_{ij}$; Z) are the Na-Cl, Na-Na, Cl-Cl
pair correlation function measured in a slice centered at $Z$,
$\rho_{\alpha}(Z)$ the average density of ion $\alpha$ near $Z$ and finally
$\lambda$ is a parameter here equal to one, but inserted for later use.

\begin{figure}\begin{center}
    \includegraphics[width=0.5\textwidth]{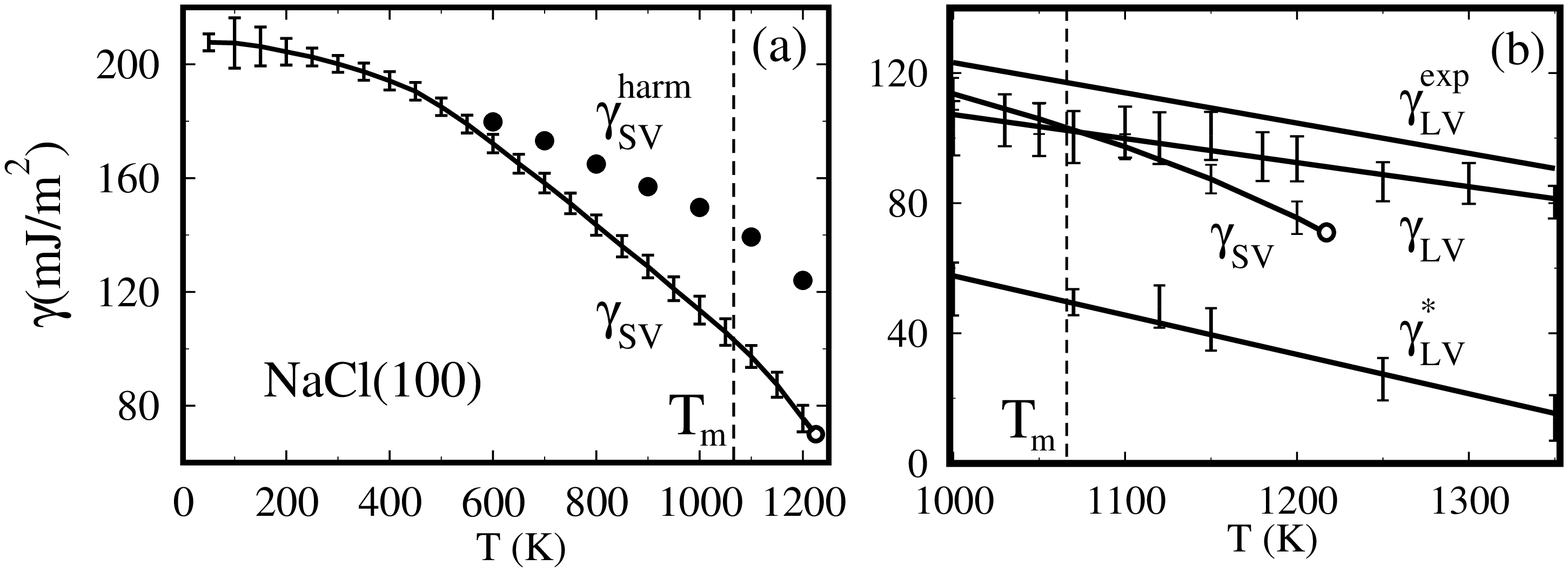}
    \caption{Calculated NaCl surface free energies.  a) Solid-vapor
    $\gamma_{SV}$.  Note the metastability up to about 150~K above
    $T_m$.  Dots: effective harmonic approximation.  b) Liquid-vapor
    $\gamma_{LV}$. Experimental data from Ref.~\cite{janz} and references
    therein.
    $\gamma_{LV}^{*}$: hypothetical liquid surface free energy calculated
    by setting $\lambda$ = 0 for outer surface atoms with coordination
    below 1.3 (highlighted in Fig.~\ref{liq}a).
    }
    \label{sva}
\end{center}\end{figure}

As shown in Fig.~\ref{sva}(b), $\gamma_{LV}$ so obtained compares well
against experiment~\cite{janz}, for example at $T_m$, $\gamma_{LV} =$
104$\pm$8 mJ/m$^2$ against 116 mJ/m$^2$.

The free energy $\gamma_{SV}$ of solid NaCl(100) was obtained
by thermodynamic integration $\gamma_{SV}/T = \gamma_{SV,0}/T_0 +
\int_{T_0}^{T} d(1/T') \Delta E(T')$, where $\Delta E(T)$ is the surface
excess internal energy obtained by simulation, and $T_{0} = $ 50~K was
chosen as a convenient reference state (our simulations are classical
and do not include quantum freezing). The results of Fig.~\ref{sva}(a)
show a relatively low $\gamma_{SV}$, characterized by a drop from 210
mJ/m$^2$ at 50~K to $\sim$ 100 mJ/m$^2$ at $T_m$. The drop confirms that
the exceptional stability of solid NaCl(100) is significantly enhanced
by anharmonicity above 600~K. To gauge the nature of anharmonicity, we
extracted from velocity-velocity correlations the $T$-dependent frequency
spectrum of a slab and of a bulk sample with same number of molecular
units. Treating both spectra as collections of harmonic oscillators,
two harmonic free energies can be calculated. Half their difference
yields an effective harmonic $\gamma^{harm}_{SV}$, whose milder drop with
temperature, due to surface vibrational softening, is seen to recover
about half the true solid surface entropy $S_{SV} = -d\gamma_{SV}/dT$. The
remaining half thus represents an additional stabilization of the solid
surface by ``hard'' anharmonicity.

Finally $\gamma_{SL}$ in the BMHFT model was calculated by inserting in
Eq.~(\ref{eq:young}) the calculated $\gamma_{SV}$, $\gamma_{LV}$ and the
contact angles $\theta$ = 50$^\circ\pm$5$^\circ$, $\phi \sim$ 0$^\circ$,
obtained by separate simulation of a liquid droplet deposited
on NaCl(100).~\cite{droplet}
We obtained in this manner $\gamma_{SL} =$~36$\pm$6~mJ/m$^2$. 
A charged hard sphere~\cite{rpm} study along the lines proposed e.g. for 
neutral hard spheres~\cite{laird} might be desirable in order to rationalize 
this result.
Our relatively large $\gamma_{SL}$ indicates poor adhesion of
the liquid to the solid substrate, attributable to the unusually large
difference of density and structure between liquid and solid.~\cite{SL}

While the result $\gamma_{SV} \simeq \gamma_{SL} + \gamma_{LV} -
40$ mJ/m$^2$ now fully accounts for the non-melting of NaCl(100) and its
incomplete triple point wetting, it does not yet clarify the role
of the liquid NaCl surface in this context.  
That can be addressed in more depth by considering the
calculated temperature dependence of the surface tension $\gamma_{LV}$
(Fig.~\ref{sva}). Strikingly, the temperature dependent drop of surface
free energy shows a factor 2.6 {\em lower} surface entropy $S_{LV}=
-d\gamma_{LV}/dT$ of the liquid surface compared with that $S_{SV}$  of
the solid surface. This is contrary to naive expectations, based on the
pictures of a relatively ordered, defect free solid surface and a very
disordered, strongly fluctuating liquid surface. The presence of this
``liquid surface entropy deficit'' (SED) suggests some underlying short
range order, that could also explain why the liquid surface tension, is here,
surprisingly, as high as that of the solid surface. The surface order,
if any, is clearly not layering: so what else could it be instead?

The answer, as was foreshadowed earlier on~\cite{goodisman}, is that
charge order, already important in bulk, plays an enhanced role at the
molecular liquid surface. If surface thermal fluctuations are indeed
large, we find them revealingly {\em correlated}. For a Na$^+$ ion
that instantaneously moves e.g., out of the surface, there is at least
one accompanying Cl$^-$, also moving out; and vice versa. So while
large surface fluctuations smear the overall liquid vapor density
profile, bridging gently between the liquid and the vapor,
(Fig.~\ref{profiles}d) the two-body correlations, described e.g., by
the the Na-Cl pair correlation function $g_{+-}(\textbf{r})$, or by its
integral, the ion coordination number $N$, drop from values typical of the
bulk liquid at $T_m$, to the nonzero value of the molecular vapor, instead
of zero as in the LJ liquid. For a more quantitative understanding, we
calculated a locally defined charge coordination number:
   \begin{equation}
     N_{\pm}(Z) = \frac{1}{2 \delta z}\int_{Z - \delta z}^{Z + \delta z}
      \left\{ dZ' \rho(Z')
       \int_{r<r_{m}} d^{3}\textbf{r}\ g_{+-} (\textbf{r};Z') \right\}
\end{equation}

where $r_m =$ 4~\AA\ corresponds to the first local
minimum of $g_{+-}(r)$, and $\zeta$ is a small interval. Starting with
the ideal $N_S = 6$ of the solid, we have $N_L=4.6$ in the bulk
liquid at $T_m$.  Moving across the liquid-vapor interface $N(Z)$ drops
continuously from 4.6 downward (Fig.~\ref{liq}b). Even if simulation
statistics is lost in the vapor, $N(Z)$ must remain and remains for
large $Z$ clearly larger than or equal to $N_V\simeq 1.3$, the value
appropriate for NaCl vapor (which at $T_m$ consists for 69\% of NaCl molecules,
and 31\% of dimers~\cite{vapor}). The larger the coordination number of
atoms in the interface region, the less their configurational entropy,
the higher the surface tension. Hence incipient molecular order could
be the reason for SED of liquid NaCl.

For a test of this idea, we repeated the surface tension calculation
of liquid NaCl by only slightly and artificially altering in
Eq.~(\ref{KBnew}) the value of correlations $g_{+-}$ or, which amounts
to the same, of the forces acting among Na and Cl for the (extremely
small) fraction of outermost surface atoms whose coordination number
$N \lesssim$ 1.3.  Since the dynamics, and thus the internal energy
remain untouched in this way, the contribution of these configurations
to the surface stress gives a direct measure of the negative surface
entropy contribution to the surface tension caused by molecular short
range order at the liquid-vapor interface.  We first identified in
simulations the surface Na and Cl ions by means of a simple algorithm. We
then extracted their mean electrostatic potential value $V_i$, and
established a roughly linear connection between atom potential and
coordination~\cite{note3,long}. Finally, we modified the Coulomb part
of the charge contribution in Eq.~(\ref{KBnew}) of surface Na$^+$ ions
in the form $\lambda = \Theta(V_{0}-V_{i})$ where $\Theta$ is the step
function and $V_0 =-6.99$~eV is the value that cancels correlations for
Na$^+$ ions with $N \leq 1.3$. Though representing an exceedingly small fraction
of the surface atoms (Fig.~\ref{liq}a), removal of
the surface stress contribution by these molecularly paired Na and Cl
ions yields a considerable surface tension drop from $\gamma_{LV}
= 104$ mJ/m$^2$ to $\gamma^*_{LV} = 53$ mJ/m$^2$ (Fig.~\ref{sva}).
The increased temperature slope $|d\gamma^*_{LV}/dT|$ exactly matches
the calculated drop from $\gamma_{LV}$ to $\gamma^*_{LV}$, confirming
that it corresponds to a purely entropic gain -- the removal of
some of the SED through cancellation of molecular surface correlations.
Since now $\gamma^*_{LV}+ \gamma_{SL} < \gamma_{SV}$, (the equivalent
correction to $\gamma_{SL}$ and $\gamma_{SV}$ is utterly negligible)
recovery of this surface entropy actually suffices to provoke 
{\em complete} instead of partial wetting of NaCl(100) at the melting point.

In conclusion we found that BMHFT potential simulations and the related
surface thermodynamics explain quantitatively the incomplete wetting of
NaCl(100) by liquid NaCl at the triple point. Three elements, namely
the exceptional anharmonic stability of the solid (100) surface, the
poor adhesion of the liquid onto the solid, and a liquid surface entropy
deficit caused by incipient molecular short range charge order, all
conspire to give rise to this phenomenon in the BMHFT model of NaCl --
and most likely also in real alkali halide surfaces.

Experimentally, it should be possible to demonstrate the overheating of
NaCl(100) and other alkali halide surfaces, perhaps in the same way as
in metals~\cite{alimetois} (though here the high vapor pressure at $T_m$
suggests using techniques not relying on ultra-high vacuum). The poor
adhesion of the molten salt onto its own solid should be detectable
in nucleation.  The short-range correlations described
at the surface of molten salts could possibly become accessible
spectroscopically. Also, if these surface correlations could for example
be altered by external means, e.g., by electric fields, the wetting
angle should change accordingly.  Finally the general possibility
that some form of short range order at the liquid-vapor interface might
affect the surface tension by reducing surface entropy could prove of
wider relevance to other compound and molecular liquid surfaces.

This work was sponsored by MIUR FIRB RBAU017S8 R004,
FIRB RBAU01LX5H, and MIUR COFIN 2003 and 2004, as well as by INFM
(Iniziativa trasversale calcolo parallelo). We acknowledge illuminating
discussions with E. A. Jagla and A. C. Levi, and the early collaboration
of W. Sekkal.




\begin{thebibliography}{30}
\bibitem{note1}
    Evaporation of molecules and molecular dimers, an event too rare to
    take place efficiently in simulations, does not appreciably alter
    the solid surface behavior described, which applies to the large flat
    solid terraces in sublimation equilibrium with the vapor at and above
    $T_m$. Similarly in liquid surface simulations the vapor is essentially
    absent, but the properties derived should be indistinguishable from
    those of the fully equilibrated liquid-vapor interface.
\bibitem{note2}
    Capillary wave induced smearing of the profile was checked and
    found to be practically irrelevant at our small sizes.~\cite{long}
    See however Ref.~\cite{aguado}.
\bibitem{note3}
    The value of $-V_{0}$ at Na is 6.80 eV in the NaCl simulated molecule,
    7.485 eV in  Na$_{2}$Cl$_{2}$, 7.82 eV in Na$_{3}$Cl$_{3}$, 8.32
    eV for liquid NaCl, 8.57 eV for solid NaCl both at $T_m$.
\end{thebibliography}
\end{document}